# $\aleph_0$ - Extended Supersymmetric Chern-Simons Theory for Arbitrary Gauge Groups [1]

Hitoshi NISHINO[2] and Subhash RAJPOOT[3]

*Department of Physics & Astronomy*
*California State University*
*1250 Bellflower Boulevard*
*Long Beach, CA 90840*

## Abstract

We present a model of supersymmetric non-Abelian Chern-Simons theories in three-dimensions with arbitrarily many supersymmetries, called $\aleph_0$-extended supersymmetry. The number of supersymmetry $N$ equals the dimensionality of any non-Abelian gauge group $G$ as $N = \dim G$. Due to the supersymmetry parameter in the adjoint representation of a local gauge group $G$, supersymmetry has to be local. The minimal coupling constant is to be quantized, when the homotopy mapping is nontrivial: $\pi_3(G) = \mathbb{Z}$. Our results indicate that there is still a lot of freedom to be explored for Chern-Simons type theories in three dimensions, possibly related to M-theory.

PACS: 11.30.Pb, 11.15.Bt, 11.15.q, 04.65.+e, 11.25.Tq
Key Words: Chern-Simons, Alephnull Extended Supersymmetry, Supergravity, Topology

---

[1] Work supported in part by NSF Grant # 0308246
[2] E-Mail: hnishino@csulb.edu
[3] E-Mail: rajpoot@csulb.edu

## 1. Introduction

Three-dimensional space-time (3D) is peculiar in the sense that Chern-Simons (CS) theories [1][2][3] can accommodate arbitrarily many supersymmetries [4][5][6][7]. Some typical examples were given in [4] with the gauge groups $OSp(p|2;\mathbb{R}) \otimes OSp(q|2;\mathbb{R})$, or with $N=8M$ and $N=8M-2$ supersymmetries in [5], or $SO(N)$ symmetries in [6]. Another example is conformal supergravity CS theory with $^\forall N$-extended supersymmetries with local $SO(N)$ gauge symmetries [8][5]. Since arbitrarily many supersymmetries are allowed in these models [4][5][6][7][8], we sometimes call them $\aleph_0$-extended supersymmetric models.

However, all of these known theories so far have rather limited gauge groups, such as $OSp$ [4] or $SO(N)$ [5]. In this brief report, we generalize these results, constructing supersymmetric CS (SCS) theory with non-Abelian gauge field strengths for arbitrary gauge group $G$. In our formulation, $G$ can be any arbitrary classical gauge group $A_n$, $B_n$, $C_n$, $D_n$, as well as any exceptional gauge group $F_2$, $G_4$, $E_6$, $E_7$ and $E_8$. We can include nontrivial trilinear interactions for non-Abelian gauge group $G$. In our system, the number of supersymmetries equals the dimensionality of the gauge group as $N = \dim G$.

## 2. $\aleph_0$-Extended Local Supersymmetries with Arbitrary Gauge Group

Our model is based on spinor charges transforming as the adjoint representation of an arbitrary gauge group $^\forall G$. It has been well-known that local supersymmetry is needed, when spinor charges are non-singlet under a local gauge group $G$. In our model, the gauge group $G$ can be completely arbitrary with the relationship $N = \dim G$.

Our field content is similar to the models given in section 2 in [6], namely, $(e_\mu{}^m, \psi_\mu{}^I, A_\mu{}^I, B_\mu{}^I, C_\mu{}^I, \lambda^I)$ where $\psi_\mu{}^I$ is the gravitino in the adjoint representation of $^\forall G$, while both $B_\mu{}^I$ and $A_\mu{}^I$ play a role of the gauge field for $G$, while $C_\mu{}^I$ is a vector field, transforming as the adjoint representation of $G$. The gaugino field $\lambda^I$ is a Majorana spinor. Our total action $I \equiv \int d^3x \, \mathcal{L}$ has the lagrangian composed of the three main parts: a supergravity lagrangian, a $BF$-type term and SCS terms, as

$$\mathcal{L} \equiv -\tfrac{1}{4}eR(\omega) + \tfrac{1}{2}\epsilon^{\mu\nu\rho}(\overline{\psi}_\mu{}^I D_\nu \psi_\rho{}^I) + \tfrac{1}{2}g\epsilon^{\mu\nu\rho}C_\mu{}^I G_{\nu\rho}{}^I$$
$$+ \tfrac{1}{2}gh\epsilon^{\mu\nu\rho}(F_{\mu\nu}{}^I A_\rho{}^I - \tfrac{1}{3}gf^{IJK}A_\mu{}^I A_\nu{}^J A_\rho{}^K) + \tfrac{1}{2}ghe(\overline{\lambda}^I \lambda^I) \quad , \tag{2.1}$$

where we adopt the signature $(\eta_{mn}) = \text{diag.}(-,+,+)$ and

$$D_{[\mu}\psi_{\nu]}{}^I \equiv \partial_{[\mu}\psi_{\nu]}{}^I + \tfrac{1}{4}\omega_{[\mu|}{}^{rs}\gamma_{rs}\psi_{|\nu]}{}^I + gf^{IJK}B_{[\mu}{}^J \psi_{\nu]}{}^K \quad . \tag{2.2a}$$

$$F_{\mu\nu}{}^I \equiv \partial_\mu A_\nu{}^I - \partial_\nu A_\mu{}^I + gf^{IJK}A_\mu{}^J A_\nu{}^K \quad , \tag{2.2b}$$

$$G_{\mu\nu}{}^I \equiv \partial_\mu B_\nu{}^I - \partial_\nu B_\mu{}^I + g f^{IJK} B_\mu{}^J B_\nu{}^K \ , \tag{2.2c}$$

$$H_{\mu\nu}{}^I \equiv (\partial_\mu C_\nu{}^I + g f^{IJK} B_\mu{}^J C_\nu{}^K) - (\mu\leftrightarrow\nu) \equiv D_\mu C_\nu{}^I - D_\nu C_\mu{}^I \ . \tag{2.2d}$$

The structure constant $f^{IJK}$ of the gauge group $^\forall G$ plays a crucial role in our formulation. The covariant derivative $D_\mu$ has both the Lorentz connection and the minimal coupling to $B_\mu{}^I$ which is the gauge field of $G$. The constants $g$ and $h$ are *a priori* nonzero and arbitrary. As usual in supergravity [9], the Lorentz connection $\omega_\mu{}^{rs}$ is an independent variable with an algebraic field equation [10]:

$$\omega_{mrs} \doteq \widehat{\omega}_{mrs} \equiv \tfrac{1}{2}(\widehat{C}_{mrs} - \widehat{C}_{msr} + \widehat{C}_{srm}) \ , \tag{2.3a}$$

$$\widehat{C}_{\mu\nu}{}^r \equiv \partial_\mu e_\nu{}^r - \partial_\nu e_\mu{}^r - (\overline{\psi}_\mu{}^I \gamma^r \psi_\nu{}^I) \ . \tag{2.3b}$$

Note that the $CG$-term is nothing but a $BF$-term.

The structure of the lagrangian (2.1) is very similar to that in section 2 in [6], but there are also differences. The most important one is that the gauge group $G$ in the present case is completely arbitrary, and the gravitino is in the adjoint representation of $G$. There is *no* restriction on the gauge group $G$.

Our total action $I$ is invariant under local supersymmetry

$$\delta_Q e_\mu{}^m = + (\overline{\epsilon}^I \gamma^m \psi_\mu{}^I) \ , \tag{2.4a}$$

$$\begin{aligned}\delta_Q \psi_\mu{}^I &= + \partial_\mu \epsilon^I + \tfrac{1}{4}\widehat{\omega}_\mu{}^{rs}\gamma_{rs}\epsilon^I + g f^{IJK} B_\mu{}^J \epsilon^K + g f^{IJK}\gamma^\nu \epsilon^J \widehat{H}_{\mu\nu}{}^K \\ &\equiv + D_\mu \epsilon^I + g f^{IJK}\gamma^\nu \epsilon^J \widehat{H}_{\mu\nu}{}^K \ ,\end{aligned} \tag{2.4b}$$

$$\delta_Q A_\mu{}^I = + f^{IJK}(\overline{\epsilon}^J \gamma_\mu \lambda^K) \ , \tag{2.4c}$$

$$\delta_Q B_\mu{}^I = + f^{IJK}(\overline{\epsilon}^J \gamma^\nu \mathcal{R}_{\mu\nu}{}^K) + h f^{IJK}(\overline{\epsilon}^J \gamma_\mu \lambda^K) \ , \tag{2.4d}$$

$$\delta_Q C_\mu{}^I = - f^{IJK}(\overline{\epsilon}^J \psi_\mu{}^K) + h f^{IJK}(\overline{\epsilon}^J \gamma_\mu \lambda^K) \ , \tag{2.4e}$$

$$\begin{aligned}\delta_Q \lambda^I &= - \tfrac{1}{2} f^{IJK}\gamma^{\mu\nu}\epsilon^J(2F_{\mu\nu}{}^K + G_{\mu\nu}{}^K + \widehat{H}_{\mu\nu}{}^K) - \tfrac{1}{2}\lambda^I(\overline{\epsilon}^J \gamma^\mu \psi_\mu{}^J) \\ &\quad - 2 f^{IJK} f^{KLM}(\gamma_{[\mu}\psi_{\nu]}{}^J)(\overline{\epsilon}^L \gamma_\mu \widetilde{\mathcal{R}}_\nu{}^M) \ ,\end{aligned} \tag{2.4f}$$

As usual in supergravity [9], $\widehat{H}_{\mu\nu}{}^I$ is the supercovariantization of $H_{\mu\nu}{}^I$:

$$\widehat{H}_{\mu\nu}{}^I \equiv H_{\mu\nu}{}^I + f^{IJK}(\overline{\psi}_\mu{}^J \psi_\nu{}^K) \ , \tag{2.5}$$

and $\mathcal{R}_{\mu\nu}{}^I$ is the gravitino field strength:

$$\mathcal{R}_{\mu\nu}{}^I \equiv (\partial_\mu \psi_\nu{}^I + \tfrac{1}{4}\widehat{\omega}_\mu{}^{rs}\gamma_{rs}\psi_\nu{}^I + g f^{IJK} B_\mu{}^J \psi_\nu{}^K) - (\mu\leftrightarrow\nu) = D_\mu \psi_\nu{}^I - D_\nu \psi_\mu{}^I \ , \tag{2.6}$$

while $\widetilde{\mathcal{R}}_\mu{}^I$ is its Hodge dual: $\widetilde{\mathcal{R}}_m{}^I \equiv (1/2)\epsilon_m{}^{rs}\mathcal{R}_{rs}{}^I$.

The on-shell closure of our system is rather easy to see, because of the field equations

$$F_{\mu\nu}{}^I \doteq 0 , \quad G_{\mu\nu}{}^I \doteq 0 , \quad \widehat{H}_{\mu\nu}{}^I \doteq 0 , \quad \mathcal{R}_{\mu\nu}{}^I \doteq 0 , \quad \lambda^I \doteq 0 , \qquad (2.7)$$

where the symbol $\doteq$ is for a field equation. Among the field strengths $F$, $G$, $H$, only $H$ has the field equation with the supercovariantized field strength, due to the minimal $gB\psi$-coupling. To be more specific, the closure of gauge algebra is

$$[\delta_Q(\epsilon_1), \delta_Q(\epsilon_2)] = \delta_P(\xi^m) + \delta_G(\xi^m) + \delta_Q(\epsilon_3^I) + \delta_L(\lambda^{rs}) + \delta_\Lambda + \delta_{\widetilde{\Lambda}} , \qquad (2.8)$$

where $\delta_P$, $\delta_G$ and $\delta_L$ are respectively the translation, general coordinate and local Lorentz transformations, while $\delta_\Lambda$ is the gauge transformation of the group $G$, and $\delta_{\widetilde{\Lambda}}$ is an extra symmetry of $C_\mu{}^I$ for our action, acting like

$$\delta_{\widetilde{\Lambda}} C_\mu{}^I = \partial_\mu \widetilde{\Lambda}^I + g f^{IJK} B_\mu{}^J \widetilde{\Lambda}^K \equiv D_\mu \widetilde{\Lambda}^I , \qquad (2.9)$$

leaving other fields intact. The parameters in (2.8) are

$$\xi^m \equiv +(\bar{\epsilon}_2^I \gamma^m \epsilon_1^I) , \quad \lambda^{rs} \equiv +\xi^\mu \widehat{\omega}_\mu{}^{rs} + 2g f^{IJK} (\bar{\epsilon}_1^I \epsilon_2^J) \widehat{H}^{rsK} , \qquad (2.10\text{a})$$

$$\epsilon_3^I \equiv -\xi^\mu \psi_\mu{}^I , \quad \Lambda^I \equiv -\xi^\mu A_\mu{}^I , \quad \widetilde{\Lambda}^I \equiv -\xi^\mu B_\mu{}^I . \qquad (2.10\text{b})$$

Due to the field equation (2.7), the existence of the last term with $\widehat{H}$ in (2.10a) does not matter for on-shell closure.

Since the parameters $g$ and $h$ have been arbitrary, we can think of interesting cases. First, if $h = 0$, then we have no SCS terms, but with the $CG$-term which is a kind of $BF$-term. Second, if $gh \neq 0$, then we have generally some quantization for the coefficients for the CS term, when the gauge group $G$ has nontrivial $\pi_3$-homotopy mapping. To be more specific,

$$\pi_3(G) = \begin{cases} \mathbb{Z} & (\text{for } G = A_n,\ B_n,\ C_n,\ D_n\ (n \geq 2,\ G \neq D_2),\ G_2,\ F_4,\ E_6,\ E_7,\ E_8) , \\ \mathbb{Z} \oplus \mathbb{Z} & (\text{for } G = SO(4)) , \\ 0 & (\text{for } G = U(1)) . \end{cases} \qquad (2.11)$$

For a gauge group with $\pi_3(G) = \mathbb{Z}$, the quantization condition is [1]

$$gh = \frac{\ell}{8\pi} \quad (\ell = \pm 1, \pm 2, \cdots) . \qquad (2.12)$$

Therefore, as long as $h \neq 0$, the minimal coupling constant $g$ should be generally quantized in this model. Third, when we keep $gh \neq 0$ restricted as in (2.12), and take a special limit $g \to 0$, we still have the effect of the CS term leaving the action topologically nontrivial, even though we lose the minimal coupling.

## 3. Comments

In this brief report, we have presented a model of $\aleph_0$-extended supersymmetric non-Abelian CS theories. The total action is invariant under $N = \dim G$-extended and local non-Abelian gauge symmetry, closing gauge algebra. Interestingly, we have two different constants $g$ and $h$ which are *a priori* arbitrary. Depending on the gauge group $G$ with nontrivial $\pi_3$-homotopy mapping, the combination $gh$ is to be quantized as $gh = \ell/(8\pi)$ $(\ell \in \mathbb{Z})$.

The generalization of supersymmetric CS theories to certain special non-Abelian gauge group is not so surprising, like some examples for $SO(N)$ for $\forall N = 1, 2, \cdots$ shown in [6]. However, the important new aspect of our present results is that non-Abelian CS theory with arbitrarily many extended supersymmetries $\forall N$ for an arbitrary gauge group $\forall G$, to our knowledge, has been presented in this paper for the first time. Note also that it is due to the special topological feature of CS theories in 3D that makes it possible to generalize the number of supersymmetries up to infinity, consistent also with local supersymmetry.

As is usual with non-Abelian CS theories [1][2][4][5][6], our action is nontrivial, even if $F_{\mu\nu}{}^I \doteq 0$ on-shell. This is due to the presence of the non-vanishing $gA \wedge A \wedge A$-term, even for a pure-gauge solution $F_{\mu\nu}{}^I \doteq 0$. It is also important that we have nontrivial trilinear interactions for the topological vector field $A_\mu$ with arbitrarily many supersymmetries. For example, even though a prototype *free* system with $\forall N$ supersymmetries was given in [10], no generalization to trilinear interactions had been successful, *except* those in [4][5][6][7][8].

It has been commonly believed that there is a limit for $N$ in 3D for interacting models with physical fields, like the limit $N \leq 8$ in 4D [11]. In this sense, our model establishes a counter-example of such wisdom. However, our result does not seem to contradict with the general analysis on supersymmetry algebra in 3D [11]. We understand that our result is a consequence of the special feature of 3D that has not been well emphasized in the past, even though some non-interacting models in [10] had certain indication of $\aleph_0$-supersymmetry. As a matter of fact, from a certain viewpoint, the existence of $\aleph_0$-supersymmetric interacting theories in dimensions $D \leq 3$ is not unusual. For example, in 1D there are analogous $\aleph_0$-supersymmetries [12].

Physics in 3D is supposed to be closely related to M-theory, in terms of supermembrane theory [13]. Our result here seems to indicate there is still a lot of freedom to be explored for Chern-Simons theories in 3D. In fact, we have in our recent paper [14] that Chern-Simons terms with quantization arise in supermembrane action [13] upon compactifications with Killing vectors.

With these encouraging results at hand, we expect that there would be more un-explored models with $\aleph_0$-supersymmetry in 3D. We are grateful to S.J. Gates, Jr. and C. Vafa for stimulating discussions.